\documentstyle[12pt]{article}

\def\baselinestretch{1.2}
\def\href#1#2{#2}  

\newcommand{\norm}[1]{\raise.3ex\hbox{:} #1 \raise.3ex\hbox{:}\,}

\newfont{\Bbb}{msbm10 scaled 1200}     
\newcommand{\mathbb}[1]{\mbox{\Bbb #1}}
\def\IR{{\mathbb R}}
\def\IZ{{\mathbb Z}}

\def\cN{{\cal N}}

\newcommand{\beq}{\begin{equation}}
\newcommand{\eeq}{\end{equation}}
\newcommand{\beqar}{\begin{eqnarray}}
\newcommand{\eeqar}{\end{eqnarray}}
\def\appendix{{\newpage\section*{Appendix}}\let\appendix\section%
        {\setcounter{section}{0}
        \gdef\thesection{\Alph{section}}}\section}

\newcommand{\be}{\begin{equation}}
\newcommand{\ee}{\end{equation}}
\newcommand{\eel}[1]{\label{#1}\end{equation}}
\newcommand{\bea}{\begin{eqnarray}}
\newcommand{\eea}{\end{eqnarray}}
\newcommand{\eeal}[1]{\label{#1}\end{eqnarray}}
\newcommand{\baq}{\begin{equation}\begin{array}{rcl}}
\newcommand{\eaq}{\end{aryray}\end{equation}}
\newcommand{\eaql}[1]{\end{array}\label{#1}\end{equation}}
\newcommand{\beac}{\begin{equation}\begin{array}{rcl}}
\newcommand{\eeacn}[1]{\end{array}\label{#1}\end{equation}}
\newcommand{\ba}{\begin{array}}
\newcommand{\ea}{\end{array}}
\newcommand{\equ}[1]{(\ref{#1})}

\newcommand{\journal}[4]{{\rm #1~}{\bf #2}\,(#3)\,#4}

\newcommand{\ijmp}{\journal {Int. J. Mod. Phys.}}

\newcommand{\pr}{\journal {Phys. Rev.}}

\newcommand{\prl}{\journal {Phys. Rev. Lett.}}

\newcommand{\rmp}{\journal {Rev. Mod. Phys.}}

\newcommand{\cmp}{\journal {Comm. Math. Phys.}}
\newcommand{\cqg}{\journal {Class. Quantum Grav.}}

\newcommand{\np}{\journal {Nucl. Phys.}}
\newcommand{\pl}{\journal {Phys. Lett.}}
\newcommand{\mpl}{\journal {Mod. Phys. Lett.}}

\newcommand{\ptp}{\journal {Progr. Theor. Phys.}}
\newcommand{\nc}{\journal {Nuovo Cim.}}

\newcommand{\grg}{\journal {Gen. Rel. Grav.}}

\catcode`\@=11

\@addtoreset{equation}{section}
\catcode`\@=12

\def\noj#1,#2,{{\bf #1} (19#2)\ }
\def\jou#1,#2,#3,{{\sl #1\/ }{\bf #2} (19#3)\ }
\def\ann#1,#2,{{\sl Ann.\ Physics\/ }{\bf #1} (19#2)\ }
\def\cmp#1,#2,{{\sl Comm.\ Math.\ Phys.\/ }{\bf #1} (19#2)\ }
\def\ma#1,#2,{{\sl Math.\ Ann.\/ }{\bf #1} (19#2)\ }
\def\jd#1,#2,{{\sl J.\ Diff.\ Geom.\/ }{\bf #1} (19#2)\ }
\def\invm#1,#2,{{\sl Invent.\ Math.\/ }{\bf #1} (19#2)\ }
\def\cq#1,#2,{{\sl Class.\ Quantum Grav.\/ }{\bf #1} (19#2)\ }
\def\cqg#1,#2,{{\sl Class.\ Quantum Grav.\/ }{\bf #1} (19#2)\ }
\def\ijmp#1,#2,{{\sl Int.\ J.\ Mod.\ Phys.\/ }{\bf A#1} (19#2)\ }
\def\jmphy#1,#2,{{\sl J.\ Geom.\ Phys.\/ }{\bf #1} (19#2)\ }
\def\jams#1,#2,{{\sl J.\ Amer.\ Math.\ Soc.\/ }{\bf #1} (19#2)\ }
\def\grg#1,#2,{{\sl Gen.\ Rel.\ Grav.\/ }{\bf #1} (19#2)\ }
\def\mpl#1,#2,{{\sl Mod.\ Phys.\ Lett.\/ }{\bf A#1} (19#2)\ }
\def\nc#1,#2,{{\sl Nuovo Cim.\/ }{\bf #1} (19#2)\ }
\def\np#1,#2,{{\sl Nucl.\ Phys.\/ }{\bf B#1} (19#2)\ }
\def\pl#1,#2,{{\sl Phys.\ Lett.\/ }{\bf #1B} (19#2)\ }
\def\pla#1,#2,{{\sl Phys.\ Lett.\/ }{\bf #1A} (19#2)\ }
\def\pr#1,#2,{{\sl Phys.\ Rev.\/ }{\bf #1} (19#2)\ }
\def\prd#1,#2,{{\sl Phys.\ Rev.\/ }{\bf D#1} (19#2)\ }
\def\prl#1,#2,{{\sl Phys.\ Rev.\ Lett.\/ }{\bf #1} (19#2)\ }
\def\prp#1,#2,{{\sl Phys.\ Rept.\/ }{\bf #1C} (19#2)\ }
\def\ptp#1,#2,{{\sl Prog.\ Theor.\ Phys.\/ }{\bf #1} (19#2)\ }
\def\ptpsup#1,#2,{{\sl Prog.\ Theor.\ Phys.\/ Suppl.\/ }{\bf #1}
(19#2)\ }
\def\rmp#1,#2,{{\sl Rev.\ Mod.\ Phys.\/ }{\bf #1} (19#2)\ }
\def\yadfiz#1,#2,#3[#4,#5]{{\sl Yad.\ Fiz.\/ }{\bf #1} (19#2) #3%
\ [{\sl Sov.\ J.\ Nucl.\ Phys.\/ }{\bf #4} (19#2) #5]}
\def\zh#1,#2,#3[#4,#5]{{\sl Zh..\ Exp.\ Theor.\ Fiz.\/ }{\bf #1}
(19#2) #3%
\ [{\sl Sov.\ Phys.\ JETP\/ }{\bf #4} (19#2) #5]}

\begin{document}

\begin{titlepage}

\begin{flushright}
CERN-TH/99-135\\
hep-th/9905148
\end{flushright}
\vfil\vfil

\begin{center}

{\Large {\bf The D4-D8 Brane System and \\
Five Dimensional Fixed Points}}

\vfil

Andreas Brandhuber  and Yaron Oz

\vfil

Theory Division, CERN\\
CH-1211, Geneva 23, Switzerland

\end{center}

\vspace{5mm}

\begin{abstract}
\noindent 
We construct dual Type I' string descriptions to five dimensional 
supersymmetric fixed points with $E_{N_f+1}$ global symmetry. 
The background is obtained as the near horizon geometry of the D4-D8 
brane system in massive Type IIA supergravity. 
We use the dual description to deduce some properties of the fixed points.

\end{abstract}

\vfil\vfil\vfil
\begin{flushleft}
May 1999
\end{flushleft}
\end{titlepage}

\newpage
\renewcommand{\baselinestretch}{1.05}  

\section{Introduction}

The consideration of the near horizon geometry of branes on one hand, and
the low energy dynamics on their worldvolume on the other hand 
has lead to conjectured duality relations between field  theories and 
string theory (M theory) on certain backgrounds \cite{mal,GKP98,WittenAdS} 
\footnote{For a recent review of the subject see \cite{us}.}.
The field theories under discussion are in various dimensions,
can be conformal or not, and with or without supersymmetry.
These properties are reflected by the type of string/M theory 
backgrounds of the dual description.

In this note we will construct dual 
Type I' string descriptions to five dimensional supersymmetric
fixed points with $E_{N_f+1}$  global symmetry\footnote{$E_{N_f+1} =
(E_8,E_7,E_6,E_5 = Spin(10),E_4=SU(5),
E_3=SU(3) \times SU(2), E_2=SU(2) \times U(1), 
E_1 = SU(2))$.}. 
These fixed points are obtained in the limit of infinite bare coupling
of $\cN=2$ supersymmetric gauge theories with gauge
group $Sp(Q_4)$, $N_f < 8$ massless hypermultiplets in the 
fundamental representation and one
massless hypermultiplet in the anti-symmetric representation 
\cite{Seiberg96,IMS97}. These theories were studied in the context of 
the AdS/SCFT correspondence in \cite{Ferrara}.
The dual background is obtained 
as the near horizon geometry of the D4-D8 brane system in massive 
Type IIA supergravity. 
The ten dimensional space 
is a fibration of $AdS_6$ over $S^4$ and has the isometry group  
$SO(2,5) \times SO(4)$.
This space provides the spontaneous compactification of
massive Type IIA supergravity in ten dimensions to the $F(4)$ 
gauged supergravity in six dimensions \cite{Romans2}.

The paper is organized as follows. In section 2 we will discuss the 
D4-D8 brane system and its relation to the five dimensional fixed points.
In section 3 we will construct the dual string description and use it
to deduce some properties of the fixed points.

\section{The D4-D8 Brane System}

We start with Type I string theory on $\IR^9 \times S^1$ with $N$ coinciding
D5 branes wrapping the circle.
The six dimensional D5 brane worldvolume theory possesses $\cN=1$ 
supersymmetry. It has 
an $Sp(N)$ gauge group, one hypermultiplet
in the antisymmetric representation of $Sp(N)$ from the DD sector and 
16 hypermultiplets
in the fundamental representation from the DN sector. 
Performing T-duality on the circle results in Type I' theory compactified on 
the interval $S^1/\IZ_2$ with two orientifolds
(O8 planes) located at the fixed points. The D5 branes become
D4 branes and there are 16 D8 branes located at points on 
the interval. They cancel the -16 units of D8 brane charge carried
by the two O8 planes. The locations of the D8 branes  
correspond to masses for the hypermultiplets in the fundamental
representation arising from the open strings between the D4 branes and
the D8 branes. The  hypermultiplet
in the antisymmetric representation is massless.

Consider first $N=1$, namely one D4 brane. The worldvolume
gauge group is $Sp(1) \simeq SU(2)$. The five-dimensional
vector multiplet contains as bosonic fields the gauge field
and one real scalar. The scalar parametrizes the location of the D4 branes
in the interval, and the gauge group is broken to $U(1)$ unless the 
D4 brane is located at one of the fixed points. 
A hypermultiplet contains four real (two complex) scalars.
The $N_f$ massless matter hypermultiplets in the fundamental and the 
antisymmetric
hypermultiplet (which is a trivial representation for $Sp(1)$)
parametrize the Higgs branch of the theory. 
It is the moduli space
of $SO(2N_f)$ one-instanton. The theory has an $SU(2)_R$ R-symmetry. The 
two supercharges as well as the scalars in the hypermultiplet transform 
as a doublet under $SU(2)_R$. In addition,
the theory has a global $SU(2) \times SO(2N_f) \times U(1)_I$ symmetry.
The $SU(2)$ factor of the global symmetry group is associated with the 
massless antisymmetric hypermultiplet and is only present if $N > 1$, 
the $SO(2N_f)$ group is associated with the $N_f$ massless
hypermultiplets in the fundamental and the $U(1)_I$ part corresponds 
to the instanton number conservation.

Consider the D8 brane background metric. It takes the form
\be\label{1} 
ds^2 = H_8^{-1/2} (-dt^2 + dx_1^2 + \ldots + dx_8^2) + H_8^{1/2} dz^2 ~~,
~~e^{-\phi} = H_8^{5/4} \ .
\ee
$H_8$ is a harmonic function on the interval parametrized by $z$.
Therefore $H_8$ is a piecewise linear function in $z$ where the slope
is constant between two D8 branes and decreases by one unit for each D8
brane crossed. 
Thus,
\be
H_8(z) = c + 16 \frac{z}{l_s} - \sum_{i=1}^{16} \frac{|z - z_i|}{l_s} - 
\sum_{i=1}^{16} \frac{|z + z_i|}{l_s} \ ,
\eel{d8backg}
where the $z_i$ denote the locations of the 16 D8 branes.

Denote the D4 brane worldvolume coordinates by $t,x_1 \ldots x_4$.
The D4 brane is
located at some point in $x_5\ldots x_8$ and $z$. 
We can consider it to be a probe of the D8 branes background.
The gauge coupling $g$  of the D4 brane worldvolume theory
and the harmonic function
$H_8$ are related as can be seen 
by expanding the DBI action of a D4 brane in the background \equ{d8backg}.
We  get 
\be\label{3}
g^2=\frac{l_s}{H_8} \ ,
\ee
where $g_{cl}^2 = \frac{c}{l_s}$ corresponds to the classical gauge coupling. 
In the field theory limit we take $l_s \rightarrow 0$ keeping
the gauge coupling $g$ fixed, 
thus, 
\be\label{gym}
g^2=\mbox{fixed}, ~~\Rightarrow ~~ \phi= \frac{z}{l_s^2}=
\mbox{fixed}, ~~l_s \to 0 \ .
\ee
In this limit we have
\be
\frac{1}{g^2} = \frac{1}{g_{cl}^2} + 16 \phi - \sum_{i=1}^{16}|\phi - m_i| - 
\sum_{i=1}^{16}|\phi + m_i|\ ,
\eel{coupling}
where the masses $m_i=\frac{z_i}{l_s^2}$.
Note that in the field theory limit we are studying the region
near $z=0$. The coordinate $\phi$ takes values in $\IR^{+}$ and 
parametrizes the 
field theory Coulomb branch.

Seiberg argued \cite{Seiberg96} that the theory at the origin
of the Coulomb branch obtained in the limit $g_{cl} = \infty$ with 
$N_f < 8$ massless hypermultiplets is a non trivial fixed point.
 The restriction $N_f < 8$ can be seen in the supergravity description
as a requirement for the harmonic function
$H_8$ in equation (\ref{d8backg}) to be positive when $c=0$
and $z_i=0$.
At the fixed point the global symmetry is enhanced
to $SU(2) \times E_{N_f+1}$.
The Higgs branch is expected to become the moduli space of  
$E_{N_f+1}$ one-instanton.

The generalization to $N=Q_4$ D4 branes is straightforward \cite{IMS97}.
The gauge group is now $Sp(Q_4)$ and the global symmetry is as before.
The Higgs branch is now the moduli space of $SO(2N_f)$ $Q_4$-instantons.
At the fixed point the global symmetry is enhanced as before and 
the Higgs branch is expected to become the moduli space of  $E_{N_f+1}$ 
$Q_4$-instanton.
Our interest in this paper will be in finding dual string (supergravity)
descriptions of these fixed points.

\section{The Supergravity (String) Description}

The low energy description of our system is given by Type I' supergravity
\cite{PW}. 
The region between two D8 branes is, as discussed in  \cite{Polchinski},
described  by massive Type IIA supergravity \cite{Romans1}.  
The configurations that we will study
have $N_f$ D8 branes located at one O8 plane and $16-N_f$ D8 branes
at the other O8 plane. Therefore we are always between D8 branes and never
encounter the situation where we cross D8 branes, and the massive Type IIA 
supergravity description is sufficient.  

The bosonic part of the massive Type IIA 
action (in string frame)
including a six-form gauge field strength which is the
dual of the RR four-form field strength is
\be\label{massiveIIA}
S = \frac{1}{2 \kappa_{10}^2}\int d^{10}x \sqrt{-g}
\left( e^{-2 \Phi}\left(R + 4 \partial_\mu \Phi
\partial^\mu \Phi \right) - \frac{1}{2 \cdot 6!} |F_6|^2 
- \frac{1}{2}  m^2 \right) 
\ee
where the mass parameter is given by\footnote{We use the conventions 
$\kappa_{10} = 8 \pi^{7/2} l_s^4$ and $\mu_8 = (2\pi)^{-9/2} l_s^{-5}$
for the gravitational coupling and the D8 brane charge, respectively.}. 
\be
m = \sqrt{2} (8-N_f) \mu_8 \kappa_{10} = \frac{8-N_f}{2 \pi l_s} \ .
\label{m}
\ee

The Einstein equations derived from
\equ{massiveIIA} read (in Einstein frame metric) 
\bea\label{eom}
2 R_{ij} &=& g_{ij}(R - \frac{1}{2} |\partial\Phi|^2  -
\frac{e^{-\Phi/2}}{2 \cdot 6!} |F_6|^2 - \frac{m^2}{2}
e^{5\Phi/2}) + \nonumber \\ 
&&+\partial_i \Phi \partial_j \Phi + \frac{e^{-\Phi/2}}{5!}
F_{im_1 \ldots m_5} F_j^{m_1 \ldots m_5} \ , \nonumber \\
0 &=&\nabla^j \partial_j \Phi - \frac{5 m^2}{4} e^{5\Phi/2} +
\frac{e^{-\Phi/2}}{4 \cdot 6!} |F_6|^2  \ , \nonumber \\ 
0 &=& \nabla^i \left( e^{-\Phi/2} F_{i m_1 \ldots m_5} \right) \ .
\eea
Except for these formulas we will be using the string frame only.

For identifying the solutions of D4 branes localized on D8 branes 
it is more convenient to start with the conformally flat form 
of the D8 brane supergravity solution.
It takes the form  \cite{PW}
\bea
ds^2 &=& \Omega(z)^2 \left( -dt^2 + \ldots + dx_4^2 + d\tilde{r}^2 +
\tilde{r}^2 d\Omega_3^2 + dz^2 \right) \ , \\
e^\Phi &=& C \left( \frac{3}{2} C m z \right)^{-\frac{5}{6}}~~,~~
\Omega(z) = \left( \frac{3}{2} C m z \right)^{-\frac{1}{6}} \ .
\nonumber
\eea
In these coordinates the harmonic function of $Q_4$ localized D4 branes 
in the near horizon limit derived from \equ{eom} reads
\be
H_4 = \frac{Q_4}{l_s^{10/3}(\tilde{r}^2 + z^2)^{5/3}} \ .
\label{H4}
\ee
This  localized  D4-D8 brane system  solution, in a different coordinate
system,  has been constructed in \cite{Youm}. 
One way to determine the harmonic function (\ref{H4}) of the localized D4 
branes is to solve  the Laplace equation in the background of the D8 branes.

It is useful to make a change of coordinates $z = r \sin \alpha, 
\tilde{r}= r \cos \alpha, 0 \leq \alpha \leq \pi/2$.
We get 
\bea\label{sol}
ds^2 &=& \Omega^2 \left( H_4^{-\frac{1}{2}}(-dt^2 + 
\ldots dx_4^2) + H_4^{\frac{1}{2}} (dr^2 + r^2 d\Omega_4^2 ) \right) 
\nonumber \ ,\\
e^\Phi &=& C \left( \frac{3}{2} C m r \sin \alpha \right)^{-\frac{5}{6}} 
H_4^{-\frac{1}{4}}~~,~~
\Omega = \left( \frac{3}{2} C m r \sin \alpha \right)^{-\frac{1}{6}} \ ,\\
F_{01234r} &=& \frac{1}{C} \partial_r \left( H_4^{-1} \right) \ ,
\nonumber
\eea
where $C$ is an arbitrary parameter of the solution \cite{PW} and
\be
d\Omega_4^2 = d\alpha^2 + (\cos \alpha)^2 d\Omega_3^2  \ .
\label{al}
\ee
The background (\ref{sol}) is a solution of the massive Type IIA supergravity 
equations (\ref{eom}).

The metric (\ref{sol}) of the D4-D8 system can be simplified to
\be
ds^2 = \left( \frac{3}{2} C m \sin \alpha\right)^{-\frac{1}{3}} 
\left( Q_4^{-\frac{1}{2}} r^\frac{4}{3} dx_{\|}^2 + Q_4^\frac{1}{2}
\frac{dr^2}{r^2} + Q_4^\frac{1}{2} d\Omega_4^2 \right) \ 
\ee
where $dx_{\|}^2 \equiv  -dt^2 + \ldots + dx_4^2$.
Define now the energy coordinate $U$ by $r^2 = l_s^5 U^3$.
That this is the energy coordinate can be seen by calculating the energy
of a fundamental string stretched in the $r$ direction or by using
the DBI action as in the previous section. 
In the field theory limit, $l_s \to 0$ with the energy $U$ fixed, 
we get the metric in the form of a {\it warped product} 
\cite{vanNieuwenhuizen} of $AdS_6 \times S^4$
\be
ds^2  = l_s^2 \left( \frac{3}{4 \pi} C (8-N_f) \sin \alpha 
\right)^{-\frac{1}{3}} 
\left( Q_4^{-\frac{1}{2}} U^2 dx_{\|}^2 + Q_4^\frac{1}{2}
\frac{9dU^2}{4U^2} + Q_4^\frac{1}{2} d\Omega_4^2 \right) \ ,
\label{warped}
\ee
and the dilaton is given by
\be
e^\Phi = Q_4^{-\frac{1}{4}} C \left(
\frac{3}{4 \pi} C (8-N_f) \sin \alpha \right)^{-\frac{5}{6}} \ .
\label{dil}
\ee

The ten dimensional space described by (\ref{warped}) is a fibration of 
$AdS_6$ over $S^4$.
It is the most general form of a metric that has the isometry of an
$AdS_6$ space \cite{van}.
The space has a boundary at $\alpha =0$ which corresponds to the location 
of the O8 plane ($z=0$).
The boundary is of the form $AdS_6 \times S^3$. 
In addition to the $SO(2,5)$ $AdS_6$ isometries, the ten dimensional 
space has also $SO(4)$ isometries associated with the spherical 
part of the metric (\ref{warped}).
In general $S^4$ has the $SO(5)$ isometry group. However, this is 
reduced due to the warped product structure. As is easily seen   
from the form of the spherical part (\ref{al}),
only transformations excluding the $\alpha$ coordinate are isometries of 
(\ref{warped}). 
We are left with an $SO(4) \sim SU(2) \times SU(2)$ isometry group. 

The two different viewpoints of the D4-D8 brane system, 
the near horizon geometry of the brane system on one hand, and
the low energy dynamics on the D4 branes worldvolume on the other hand 
suggest a duality relation. Namely,
{\it Type I' string theory compactified on the background (\ref{warped}), 
(\ref{dil}) with a 4-form flux 
of $Q_4$ units on $S^4$ is dual to an $\cN=2$ supersymmetric 
five dimensional fixed point} \footnote{The 4-form is the dual
of the 6-form in \equ{sol}.}.
The fixed point is obtained in the limit of
infinite coupling of $Sp(Q_4)$ gauge theory 
with $N_f$ hypermultiplets in the fundamental representation
and one hypermultiplet in the antisymmetric representation, where 
$m \sim (8-N_f)$ as in (\ref{m}).
The  $SO(2,5)$ symmetry of the compactification
corresponds to the conformal symmetry of the field theory.
The $SU(2) \times SU(2)$ symmetry of the compactification
corresponds $SU(2)_R$ R-symmetry and to the $SU(2)$ global symmetry
associated with the massless hypermultiplet in the antisymmetric
representation.

At the boundary $\alpha=0$ the dilaton (\ref{dil}) blows up and Type I'
is strongly coupled. In the weakly coupled dual heterotic string description
this is seen as an enhancement of the gauge symmetry to $E_{N_f+1}$.
One can see this enhancement of the gauge symmetry in the Type I' 
description by analysing the D0 brane dynamics near the orientifold plane
\cite{Bergman,Nilles, Schwimmer}.
This means that we have  $E_{N_f+1}$ vector fields that propagate on the 
$AdS_6 \times S^3$ boundary,
as in the Horava-Witten picture \cite{HW}.
The scalar curvature of the background (\ref{warped}), (\ref{dil}) 
\be
{\mathcal{R}} l_s^2 \sim (C (8-N_f))^\frac{1}{3} Q_4^{-\frac{1}{2}} 
(\sin \alpha)^{-\frac{5}{3}}  \ ,
\label{R}
\ee
blows up at the boundary as well. 
In the dual heterotic description the dilaton is small but the curvature 
is large, too.
For large $Q_4$ there is a region, $\sin \alpha \gg  Q_4^{-\frac{3}{10}}$,
where both curvature (\ref{R}) and
dilaton (\ref{dil}) are small and thus we can trust supergravity.

The $AdS_6$ supergroup is $F(4)$. Its bosonic subgroup is 
$SO(2,5) \times SU(2)$.
Romans constructed an $\cN=4$ six dimensional gauged supergravity with 
gauge group $SU(2)$ that realizes $F(4)$ \cite{Romans2}.
It was conjectured in \cite{Ferrara} that it is related to a 
compactification of the ten dimensional massive Type IIA supergravity.
Indeed, we find that the ten dimensional background space is the warped 
product of $AdS_6$ and $S^4$ (\ref{warped}) (with $N_f=0$). 
The reduction to six dimensions can be done in two steps.
First we can integrate over the coordinate $\alpha$. 
This yields a nine dimensional space of the form $AdS_6 \times S^3$.
We can then reduce on $S^3$ to six dimensions, while gauging its isometry 
group.
Roman's construction is based on gauging an $SU(2)$ subgroup of the 
$SO(4)$ isometry group.

The massive Type IIA supergravity action in the string frame
goes like 
\be
l_s^{-8} \int \sqrt{-g} e^{-2 \Phi} {\mathcal{R}} \sim Q_4^{5/2} \ ,
\ee
suggesting that the number 
of degrees of freedom goes like  $Q_4^{5/2}$ in the regime where it is 
an applicable description. Terms in the Type I' action coming from the 
D8 brane DBI action turn out to be of the same order.
Viewed from M theory point of view, we expect the corrections
to the supergravity action to go like 
$l_p^3 \sim l_s^3 e^\Phi \sim 1/Q_4$, where $l_p$ is the eleven-dimensional
Planck length. This seems to suggest that the field theory has a $1/Q_4$ 
expansion at large $Q_4$. For example, the one-loop correction
of the form 
\be
l_s^{-8} \int \sqrt{-g} l_s^6 {\mathcal{R}}^4 \sim Q_4^{1/2}
\ee
is suppressed by $Q_4^2$ compared to the tree-level action.

According to the  $AdS$/CFT correspondence the spectrum of chiral primary 
operators of the fixed point theory can be derived from the spectrum of 
Kaluza-Klein excitations of massive Type IIA supergravity on the 
background (\ref{warped}).
We will not carry out the detailed analysis here but make a few comments.
The operators fall into representations of the $F(4)$ supergroup.
As in the case of the six dimensional $(0,1)$ fixed point with $E_8$ global 
symmetry \cite{gimon} we expect $E_{N_f+1}$ neutral operators to match the 
Kaluza-Klein reduction of fields in the bulk geometry, and 
$E_{N_f+1}$ charged  operators to match the 
Kaluza-Klein reduction of fields living on the boundary.
Among the $E_{N_f+1}$ neutral operators we expect to have dimension $3k/2$
operators of the type $Tr \phi^k$ where $\phi$ is a
complex scalar in the hypermultiplet, which parametrize the Higgs branch 
of the theory.
Like in \cite{gimon} we do not expect all these operators to be in 
short multiplets.
We expect that those in long multiplets will generically receive $1/Q_4$
corrections to their anomalous dimensions.
Unlike the hypermultiplet, the vector multiplet in five dimensions is 
not a representation of the superconformal group $F(4)$. 
Therefore, we do not expect Kaluza-Klein 
excitations corresponding to neutral operators of the type 
$Tr \varphi^k$ where $\varphi$ is a real scalar in the vector multiplet,
which parametrizes the Coulomb branch of the theory. 
Among the  $E_{N_f+1}$ charged  operators we expect to have the 
dimension four $E_{N_f+1}$ global symmetry currents that couple to the 
massless $E_{N_f+1}$ gauge fields on the boundary.

\section*{Acknowledgement}
We would like to thank N. Itzhaki for participating in the early stages of this
work and for many useful discussions.
We would also like to thank A. Zaffaroni for a useful discussion.

\newpage

\begingroup\raggedright\endgroup


\begin{thebibliography}{10}



\bibitem{mal}
J.~M. Maldacena, ``The Large $N$ Limit of Superconformal Field Theories and
Supergravity,'' \href{http://xxx.lanl.gov/abs/hep-th/9711200}{{\tt
hep-th/9711200}}.

\bibitem{GKP98}
S.~S. Gubser, I.~R. Klebanov, and A.~M. Polyakov, ``Gauge theory correlators
from noncritical string theory,'' {\em Phys. Lett.} {\bf B428} (1998) 105,
\href{http://xxx.lanl.gov/abs/hep-th/9802109}{{\tt hep-th/9802109}}.

\bibitem{WittenAdS}
E.~Witten, ``Anti-de Sitter space and holography,''
\href{http://xxx.lanl.gov/abs/hep-th/9802150}{{\tt hep-th/9802150}}.

\bibitem{us} 
O. Aharony, S. S. Gubser, J. Maldacena, H. Ooguri and Y. Oz,
``Large N Field Theories, String Theory and Gravity,''
hep-th/9905111.

\bibitem{Seiberg96}
N. Seiberg, ``Five Dimensional SUSY Field Theories, Non-trivial Fixed 
Points and String 
Dynamics,'' {\em Phys. Lett.} {\bf B388} (1996) 753,
\href{http://xxx.lanl.gov/abs/hep-th/9608111}{{\tt hep-th/9608111}}.

\bibitem{IMS97}
K. Intriligator, D.R. Morrison, and N. Seiberg, ``Five-Dimensional 
Supersymmetric Gauge 
Theories and Degenerations of Calabi-Yau Spaces,'' 
{\em Nucl. Phys.} {\bf B497} (1997) 56,
\href{http://xxx.lanl.gov/abs/hep-th/9609070}{{\tt hep-th/9609070}}.


\bibitem{Ferrara}
S. Ferrara, A. Kehagias, H. Partouche and A. Zaffaroni, 
``AdS$_6$ interpretation of 5-D superconformal field theories,'' 
{\em Phys. Lett.} {\bf B431}, 57 (1998).


\bibitem{Romans2}
L.J.~Romans,
``The F(4) Gauged Supergravity In Six-Dimensions,"
{\em Nucl. Phys.} {\bf B269}, 691 (1986).



\bibitem{PW}
J.~Polchinski and E.~Witten,
``Evidence for heterotic - type I string duality,"
{\em Nucl. Phys.} {\bf B460}, 525 (1996)
hep-th/9510169.

\bibitem{Polchinski}
J.~Polchinski,
``Dirichlet Branes and Ramond-Ramond charges,"
{\em Phys. Rev. Lett.} {\bf 75}, 4724 (1995)
hep-th/9510017.

\bibitem{Romans1}
L.J.~Romans,
``Massive N=2a Supergravity In Ten-Dimensions,"
{\em Phys. Lett.} {\bf 169B}, 374 (1986).


\bibitem{Youm}
D.~Youm,
``Localized intersecting BPS branes,"
hep-th/9902208.

\bibitem{vanNieuwenhuizen}
P. van Nieuwenhuizen and N. P. Warner,
``New compactifications of ten-dimensional and 
eleven-dimensional supergravity on manifolds which
are not direct products," 
{\em Commun. Math. Phys.} {\bf 99}, 141 (1985).

\bibitem{van}
P. van Nieuwenhuizen, Les Houches 1983 Lectures,
B. de Witt and R. Stora eds., North-Holland.

\bibitem{Bergman}
O. Bergman, M. R. Gaberdiel and  G. Lifschytz,
``String Creation and Heterotic-Type I' Duality,''
{\em  Nucl.Phys.} {\bf  B524}, 524 (1998).


\bibitem{Nilles}
D. Matalliotakis, H. P. Nilles and  S. Theisen, ``Matching the BPS Spectra of 
Heterotic - Type I - Type I' Strings,''
{\em Phys.Lett.} {\bf  B421}, 169  (1998),hep-th/9710247.


\bibitem{Schwimmer}
C. P. Bachas, M. B. Green and A. Schwimmer,
``$(8,0)$ Quantum mechanics and symmetry enhancement in type I' superstrings,''
{\em JHEP} {\bf 9801}, 006 (1998),
hep-th/9712086.

\bibitem{HW} P. Horava and E. Witten,
``Heterotic and Type I String Dynamics from Eleven Dimensions,'' 
{\em Nucl.Phys.} {\bf  B460}, 
506 (1996), hep-th/9510209 ; ``Eleven-Dimensional Supergravity on a 
Manifold with Boundary,''
{\em  Nucl.Phys.} {\bf B475}, 
 94 (1996)  hep-th/9603142.







\bibitem{gimon} E. G. Gimon and  C. Popescu, ``The Operator Spectrum of the 
Six-dimensional $(1,0)$ Theory,''
{\em JHEP} {\bf 9904}, 018 (1999), hep-th/9901048.

\end{thebibliography}
\end{document}